\documentclass[pdftex,twocolumn,epjc3]{svjour3}          

\RequirePackage[T1]{fontenc}

\smartqed  

\RequirePackage{graphicx}
\RequirePackage{mathptmx}      
\RequirePackage{flushend}
\RequirePackage[numbers,sort&compress]{natbib}
\RequirePackage[colorlinks,citecolor=blue,urlcolor=blue,linkcolor=blue]{hyperref}
\usepackage{multirow}
\usepackage[fleqn]{amsmath}

\journalname{Eur. Phys. J. C}

\begin{document}

\title{Performance study of the separation of the full hadronic WW and ZZ events at the CEPC}

\author{Yongfeng Zhu\thanksref{addr1, addr2}
        \and
        Manqi Ruan\thanksref{e1,addr1} 
}

\thankstext{e1}{e-mail: ruanmq@ihep.ac.cn}

\institute{Institute of High Energy Physics, Beijing \label{addr1}
       \and
         University of Chinese Academy of Sciences, Beijing \label{addr2}
}

\date{Received: date / Accepted: date}

\maketitle

\begin{abstract}

The separation of the full hadronic WW and ZZ events is an important benchmark for the CEPC detector design and performance evaluation.
This separation performance is determined by the intrinsic boson mass distributions, the detector performance, and the jet confusion.  
The latter refers to the uncertainties induced by the jet clustering and pairing algorithms.
Using the CEPC baseline simulation, we demonstrate that the full hadronic WW and ZZ events can be efficiently separated.
We develop an analytic method that quantifies the impact of each component and conclude that the jet confusion dominates the separation performance. 
The impacts of the initial state radiations and the heavy flavor jets are also analyzed and confirmed to be critical for the separation performance.

\end{abstract}

\section{Introduction}
The Circular Electron Positron Collider (CEPC) is a proposed electron-positron collider with a total circumference of 100 km and two interaction points.
It will be operated at center-of-mass energies from 91 GeV to 240 GeV and produces large samples of the W, the Z, and the Higgs bosons.
Its nominal luminosity and massive boson yields are listed in Table~\ref{plain} \cite{r1}.
The CEPC can measure most of the Higgs boson properties with accuracies that exceed the ultimate precision of the HL-LHC by one order of magnitude, and also boost current precision of the Electroweak (EW) measurements by one order of magnitude. 
The CEPC can also be upgraded to a proton-proton collider with a center-of-mass energy around 100 TeV. 

\begin{table}
\centering
\caption{Running time, instantaneous and integrated luminosities at different values of the center-of-mass energy and anticipated corresponding boson yields at the CEPC. The Z boson yields of the Higgs factory and WW threshold scan operation are from the initial-state radiative return $e^{+}e^{-} \to \gamma Z$ process. The ranges of luminosities for the Z factory correspond to the two possible solenoidal magnetic fields, 3 or 2 Tesla.}
\label{plain}
\begin{tabular*}{\columnwidth}{@{\extracolsep{\fill}}cccc@{}}
\hline
 \multirow{2}{*}{Operation mode}   & \multirow{2}{*}{ Z factory}                &  WW                                                  &\multirow{2}{*}{Higgs factory}        \\
                                                       &                                                         & threshold scan                                  &                                                       \\
\hline 
$\sqrt{s}$ (GeV)                              & \multirow{1}{*}{91.2}                            &  \multirow{1}{*}{158 - 172}                  &  \multirow{1}{*}{ 240}                        \\
\hline
Running time                             &   \multirow{2}{*}{2 }                               &   \multirow{2}{*}{1}                               & \multirow{2}{*}{7}                            \\
(years)                                       &  &  &  \\
\hline
Instantaneous                           & \multirow{3}{*}{17 - 32}                        &   \multirow{3}{*}{10}                             &  \multirow{3}{*}{3}                           \\
Luminosity                                 &   &     &     \\
($10^{34} cm^{-2}s^{-1}$)          &   &     &     \\
\hline
Integrated Luminosity                & \multirow{2}{*}{8 - 16}                           &    \multirow{2}{*}{2.6}                            & \multirow{2}{*}{5.6}                       \\
($ab^{-1}$)                                &     &      & \\
\hline
Higgs yield                                &   -                                                           &     -                                                        & $10^{6}$                  \\
\hline
W yield                                     &   -                                                            &   $10^{7} $                                            & $10^{8}$                   \\
\hline
Z yield                                      & $10^{11 - 12}$                                        &    $10^{8}$                                             & $10^{8}$                  \\      
   \hline
\end{tabular*}
\end{table}

\begin{figure}
\begin{minipage}{\columnwidth}
\centering
\includegraphics[width=0.8\columnwidth]{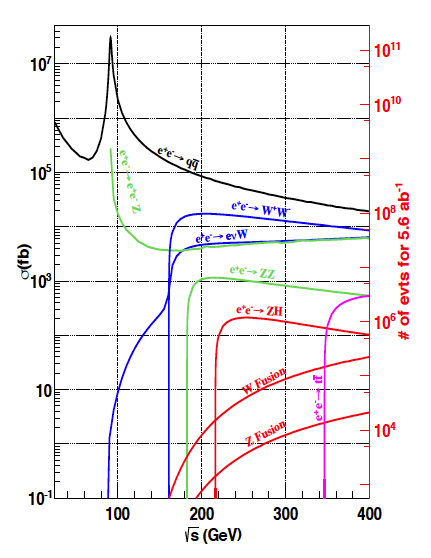}
\end{minipage}
\caption{The cross section for unpolarized $e^{+}e^{-}$ collision, the right side shows the expected number of events at the nominal parameters of the CEPC Higgs runs at 240 GeV center-of-mass energy.}
\label{crosssection}
\end{figure}

At 240 GeV center of mass energy, the Higgs boson is mainly produced through the ZH process at the CEPC. 
The leading di-boson Standard Model backgrounds for the CEPC Higgs measurements are the WW and ZZ processes, see Figure~\ref{crosssection}.
A successful separation between the Higgs signal and the di-boson backgrounds is essential for the precise Higgs measurements. 
In addition, the separation of the WW and ZZ events is important for the QCD measurement, the Triplet Gauge Boson Coupling measurement, and the W boson mass measurement at continuum.

Half of these di-boson events decay into 4-jet final states.
The separation between those 4-jet events is determined by the intrinsic boson mass distribution, the detector performance, and the jet confusion.
The latter refers to the uncertainties induced by the jet clustering and pairing algorithm.
Giving the relatively small mass difference between the W boson and the Z boson, the separation between the WW and the ZZ events in the full hadronic final states is extremely demanding in the detector performance and the jet confusion control. 
Therefore it serves as a stringent benchmark for the detector design and reconstruction algorithm development. 
Using the CEPC baseline detector geometry and software, we investigate the separation performance of the full hadronic WW and ZZ events at full simulation level.
We confirm that these events can be clearly separated with the CEPC baseline detector. 
Through comparative analyses, we quantify the impacts of each component and conclude the jet confusion dominates the separation performance.

This paper is organized as follows. 
Section 2 introduces the CEPC baseline detector geometry and the software. 
The analysis method and the separation performance at various conditions are quantified and compared in section 3. 
Using the Monte Carlo (MC) truth information, section 4 further analyzes the jet confusion. 
The conclusion is summarized in section \ref{secConclusion}. 

\section{Detector geometry, software, sample and analysis method}

\begin{figure}
\begin{minipage}{\columnwidth}
\centering
\includegraphics[width=\columnwidth]{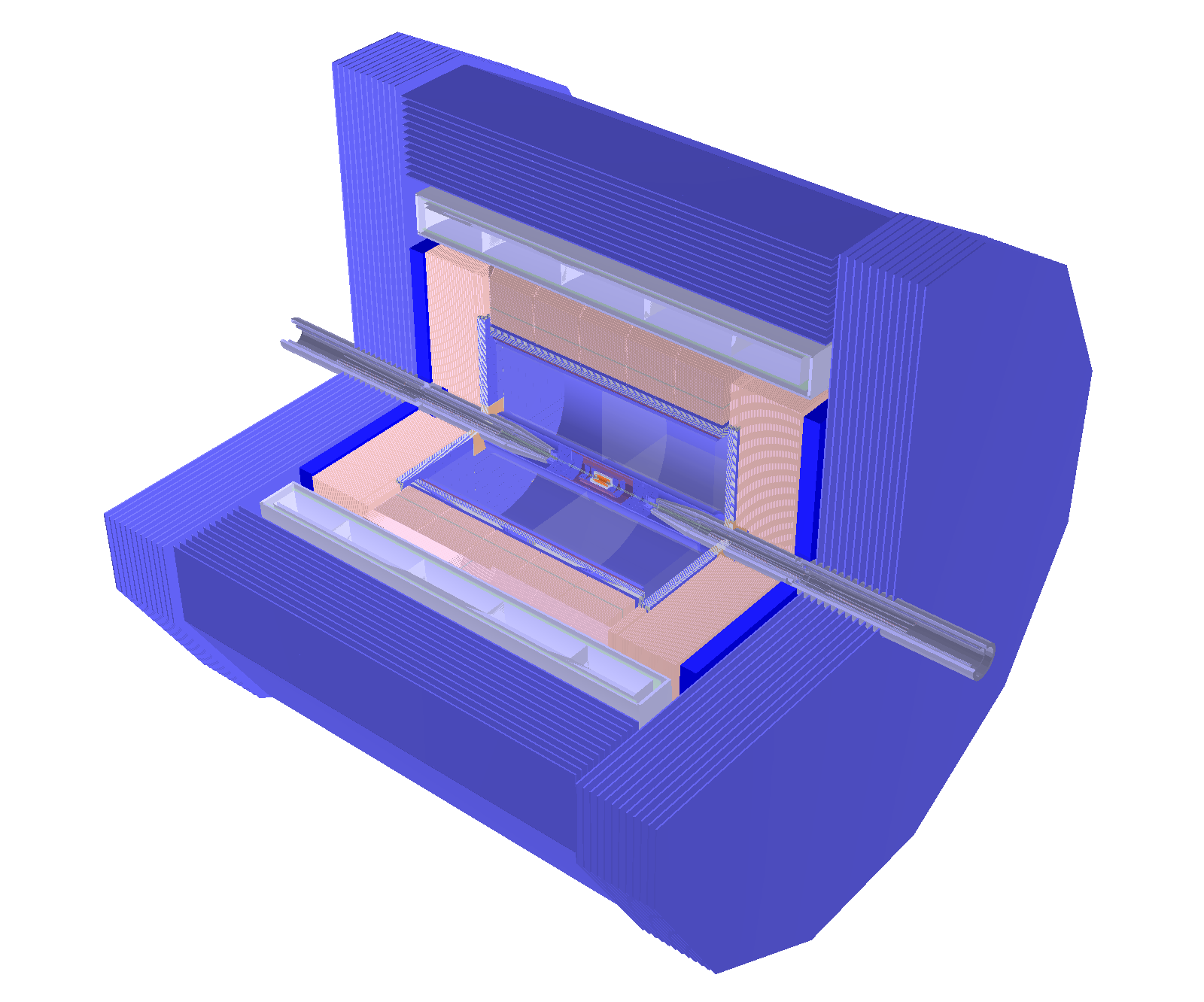}
\end{minipage}
\caption{The CEPC baseline detector. From inner to outer, the detector is composed of a silicon pixel vertex detector, a silicon inner tracker, a TPC, a silicon external tracker, an ECAL, an HCAL, a solenoid of 3 Tesla and a return yoke embedded with a muon detector. In the forward regions, five pairs of silicon tracking disks are installed to enlarge the tracking acceptance.}
\label{APODISstructure}
\end{figure}

A Particle Flow oriented detector design is the baseline detector as described the CEPC CDR~\cite{r1}. 
This baseline reconstructs all the visible final state particles in the most-suited detector subsystems.
For the CEPC physics measurements, this baseline reconstructs all the core physics objects with high efficiency, high purity, and high precision~\cite{r1}\cite{r6}. 
From inner to outer, the detector is composed of a silicon pixel vertex detector, a silicon inner tracker, a Time Projection Chamber (TPC) surrounded by a silicon external tracker, a silicon-tungsten sampling Electromagnetic Calorimeter (ECAL), a steel-Glass Resistive Plate Chambers (GRPC) sampling Hadronic Calorimeter (HCAL), a 3 Tesla superconducting solenoid, and a flux return yoke embedded with a muon detector. 
The structure of the CEPC detector is shown in Figure~\ref{APODISstructure}.
In fact, the separation of vector bosons scattering processes (with $\nu\nu$WW and $\nu\nu$ZZ final states) provides a strong motivation for the Particle Flow oriented detector design~\cite{ILCTDR}\cite{CLICCDR}.

\begin{figure}
\begin{minipage}{\columnwidth}
\centering
\includegraphics[width=\columnwidth]{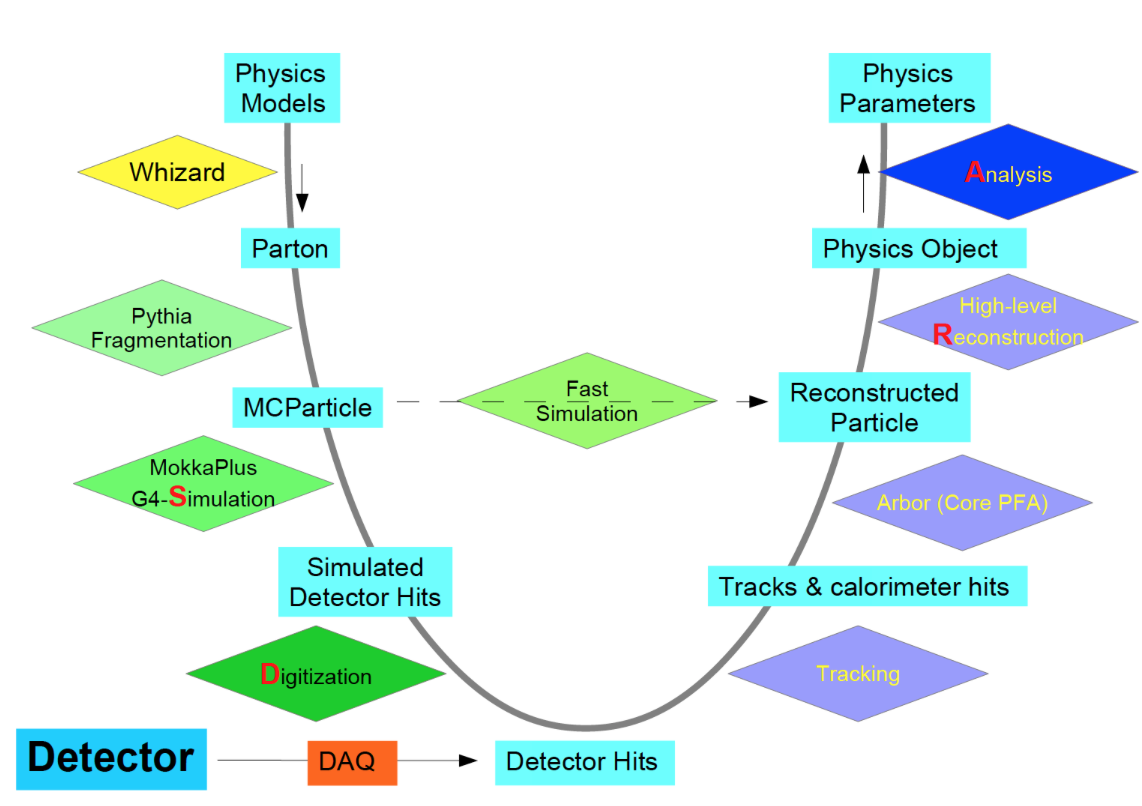}
\end{minipage}
\caption{The information flow of the CEPC software chain.}
\label{softwarechain}
\end{figure}

\begin{figure}
\begin{minipage}{\columnwidth}
\centering
\includegraphics[width=7.1cm, height = 6.8cm]{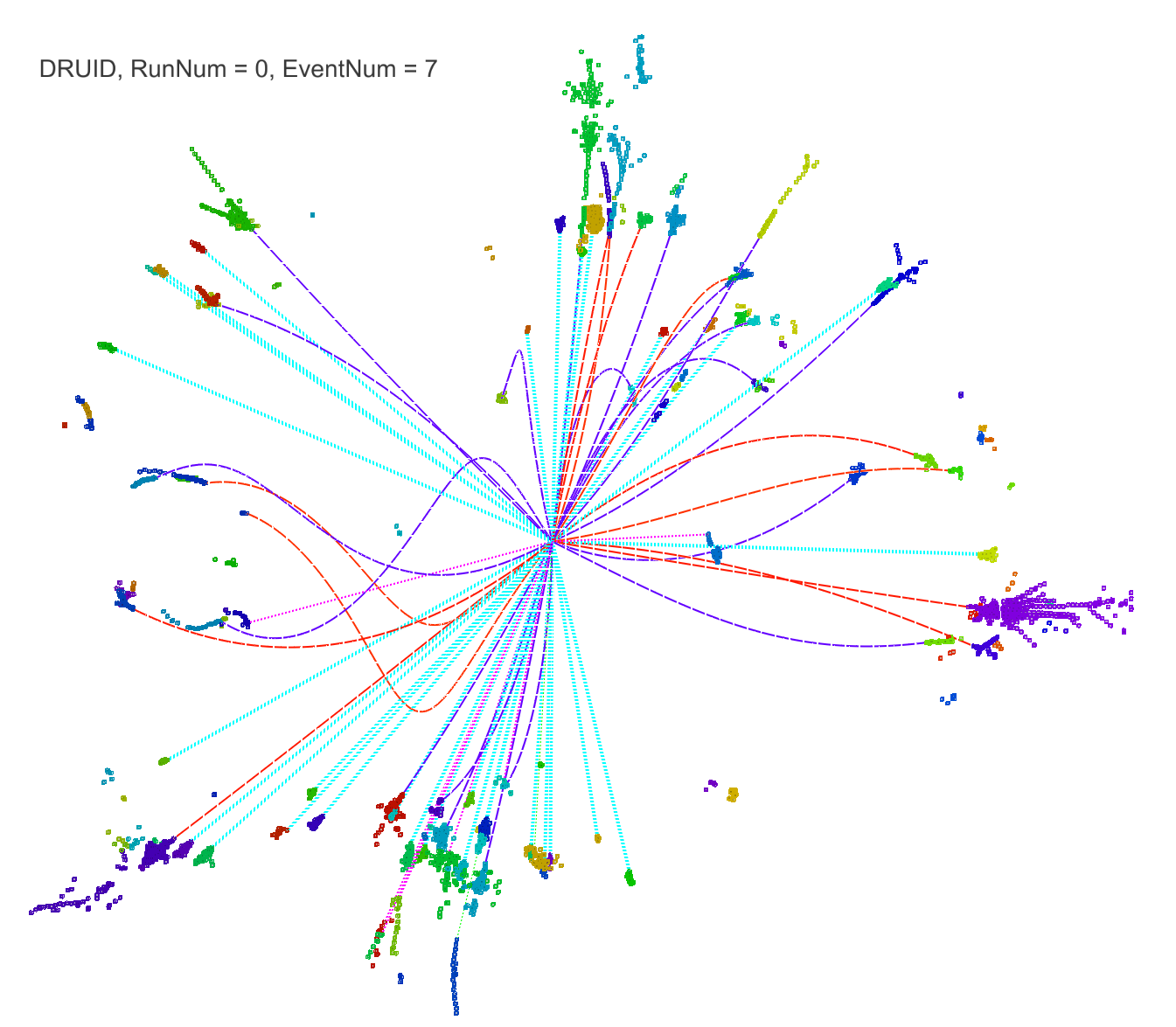}
\end{minipage}
\caption{The display of a reconstructed WW event. This event has 82 final state particles whose energy exceed 0.5 GeV, reconstructed by Arbor. The charged particles are represented by the curves (color represent their charge) associated with calorimeter clusters. The photons are displayed as cyan straight lines associated with calorimeter clusters.}
\label{Reco}
\end{figure}

The CEPC baseline software is demonstrated in Figure~\ref{softwarechain}. 
It uses the Whizard~\cite{r2}\cite{r3} and the Pythia~\cite{r4} generators as the starting point. 
The detector geometry is implemented into the MokkaPlus~\cite{r8}, a GEANT4~\cite{r5} based full simulation module.
The MokkaPlus calculates the energy deposition in the detector sensitive volumes and creates simulated hits.
For each sub-detector, the digitization module converts the simulated hits into digitized hits by convoluting the corresponding sub-detector responses. 
The reconstruction modules include the tracking, the Particle Flow, and the high-level reconstruction algorithms. 
The digitized tracker hits are reconstructed into tracks via the tracking modules. 
The particle flow algorithm, Arbor~\cite{r6}, reads the reconstructed tracks and the calorimeter hits to build reconstructed particles. 
High-level reconstruction algorithms reconstruct composite physics objects such as the converted photons, the $\tau$s, the jets, et al., and identify the flavor of the jets.

Using the CEPC baseline simulation, 
we produce inclusive samples of 38k WW and 38k ZZ events.
These samples include all the different quark flavors according to the SM decay branching ratios.
To simplify the analysis, the interference between WW and ZZ is ignored.  
To analyze the impact of heavy flavors, we also produce light flavor samples for comparison. 
These light flavor samples are 30k $WW\to u\bar{d}\bar{u}s$ or $u\bar{s}\bar{u}d$ and 27k $ZZ\to u\bar{u}u\bar{u}$ events. 
Figure~\ref{Reco} displays a reconstructed $e^{+}e^{-}\to WW \to u\bar{u}s\bar{d}$ event using Druid~\cite{r12}.
All the samples are generated at the center-of-mass energy of 240 GeV.

Starting with the fully reconstructed WW/ZZ events, our analysis employs the jet clustering and pairing algorithm.  
The reconstructed particles are clustered into four RecoJets using the $\it{k_t}$ algorithm for the $e^{+}e^{-}$ collisions ($\it{e^+e^- k_t}$) with the FastJet package~\cite{r7}. 
A minimal $\chi^{2}$ method is used for the jet pairing. 
These four RecoJets are paired into two di-jet systems. 
Their masses are compared with the hypothesis of a WW or a ZZ event via the $\chi^{2}$ defined as:

 $$\chi^{2} = \frac{(M_{12} - M_{B})^2 + (M_{34} - M_{B})^2}{\sigma_{B}^{2}}. $$ 
 
\begin{table}
\centering
\caption{The values of $\sigma_B$ for different cases.}
\label{sigma}
\begin{tabular*}{0.9\columnwidth}{@{\extracolsep{\fill}}ccc@{}}
\hline
$\sigma_B$/GeV       & $\sigma_W$     &  $\sigma_Z$          \\
\hline 
GenJet                      &   2.0                     &  2.5                         \\
\hline
RecoJet                    &   3.8                  &  4.4                          \\
\hline
\end{tabular*}
\end{table}

The quantity $M_{12}$ and $M_{34}$ refer to the masses of di-jet systems, and $M_{B}$ is the reference mass of the Z or the W boson~\cite{r9}. 
The $\sigma_{B}$ is the convolution of the boson width and the detector resolution. 
According to~\cite{r1}, the detector resolution is set to be 4\% of the boson mass. 
The values of the $\sigma_{B}$ for different cases are listed in Table~\ref{sigma}.
Among all six possible combinations (corresponding to three different jet pairings and two values of $M_{B}$), 
the one with the minimal value of the $\chi^{2}$ determines the event type and corresponding di-jet masses. 

Using the same jet clustering and pairing setup for the RecoJets analysis, the visible particles at the MC truth level can be clustered into the GenJets and paired into di-jet systems. 
These GenJets are corresponding to the perfect detector, 
and the separation performance using the GenJets describes the impacts of the intrinsic boson mass distribution and the jet confusion.
In this paper, the analyses are performed using both the RecoJets and the GenJets.

\section{Separation Performance with Overlapping Ratio } 

\begin{figure} 
\centering
\includegraphics[width=\columnwidth]{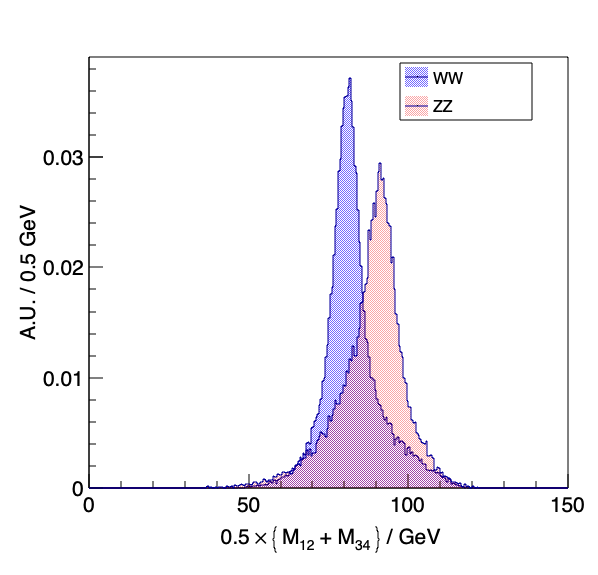}
\caption{The RecoJet level distributions of $0.5\times(M_{12} + M_{34})$ of the WW and ZZ events. The overlapping ratio is $57.8\% \pm 0.23\%$.}
\label{czwHerror}
\end{figure}

Using the method introduced above, the masses of the di-jet systems ($M_{12}$ and $M_{34}$) are calculated.  
Figure~\ref{czwHerror} shows the average reconstructed di-jet mass distributions of the inclusive WW and ZZ samples using the RecoJets, each normalized to unit area. 
Each distribution exhibits a clear peak at the anticipated boson mass and an artificial tail towards the other peak.
These tails are induced by the jet pairing algorithm, the neutrinos generated in heavy flavor quark fragmentation, and the ISR photons. 
The peaks are clearly separated, however, the tails lead to significant confusion between the WW and ZZ events. 
 
The confusion can be evaluated by the overlapping ratio between two distributions: 

$$ Overlapping\ Ratio = \sum_{bins}min({a_{i}, b_{i}}),$$

$a_{i}$ and $b_{i}$ are the bin contents of both distributions at a same bin.  
To the first order, the overlapping ratio is equal to the sum of misidentification probabilities ($P_{WW\to ZZ} + P_{ZZ \to WW}$ in this manuscript).
An overlapping ratio of zero means no mis-identification. 

Through a parameter scan of the generalised $\it{k_t}$ algorithm for the ${e^+e^-}$ collision, 
the $\it{e^+e^- k_t}$ algorithm is chosen for this analysis as it has the minimum overlapping ratio on the inclusive sample.

\begin{figure} 
\centering
\includegraphics[width=\columnwidth]{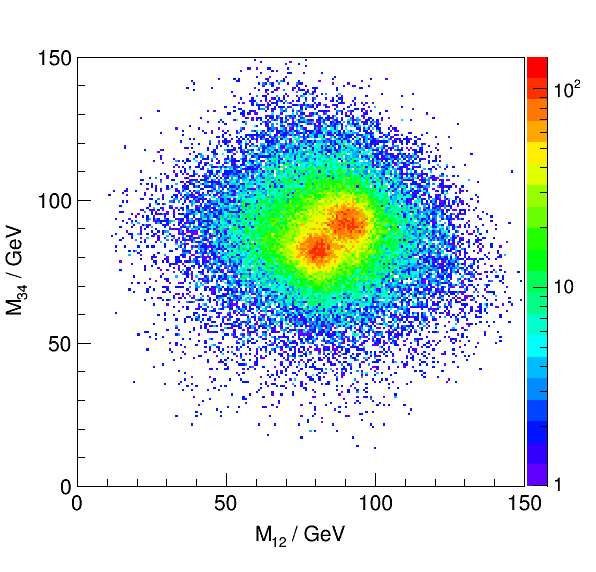}
\caption{The RecoJet level distribution of $M_{12}$ versus $M_{34}$. }
\label{czwDiserror}
\end{figure}

\begin{figure} 
\centering
\includegraphics[width=\columnwidth]{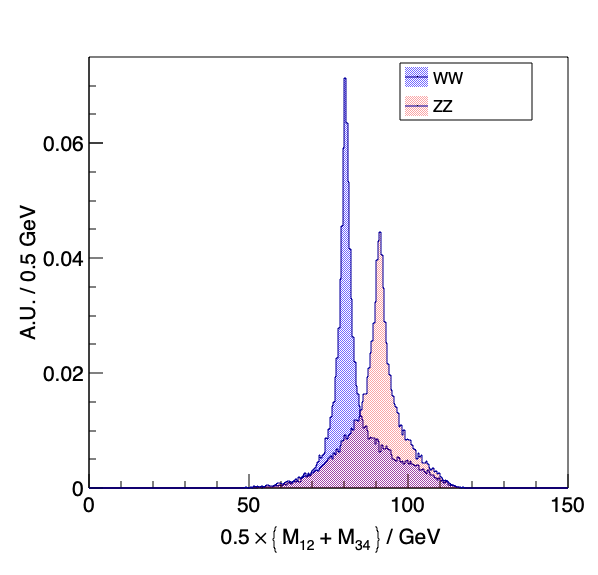}
\caption{The GenJet level distributions of $0.5\times(M_{12} + M_{34})$ of the WW and ZZ events. The overlapping ratio is $52.6\% \pm 0.25\%$.}
\label{cGzwHerror}
\end{figure}

\begin{figure} 
\centering
\includegraphics[width=\columnwidth]{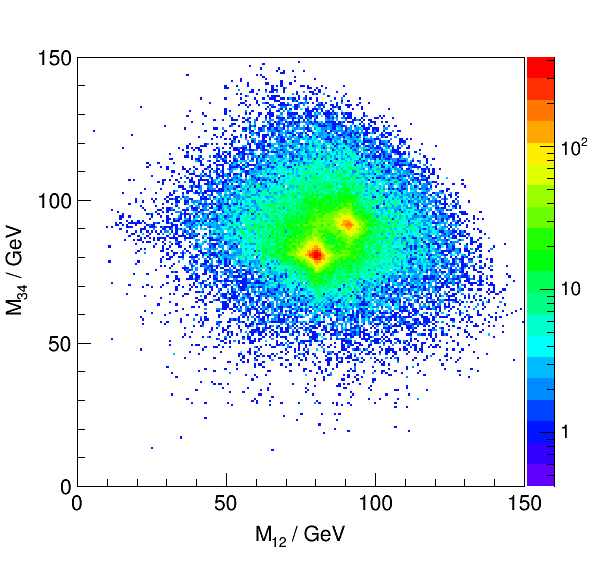}
\caption{The GenJet level distribution of $M_{12}$ versus $M_{34}$.}
\label{cGzwDiserror}
\end{figure}

Figure~\ref{czwHerror} has an overlapping ratio of $57.8\%\pm 0.23\%$.
The correlation of $M_{12}$ versus $M_{34}$ using the RecoJet is shown in Figure~\ref{czwDiserror}, the distributions of the WW and ZZ events are overlapped. 
Figure~\ref{czwDiserror} has two separable peaks located on a large area of a flat plateau. 
The latter contributes significantly to the overlapping ratio. 

The separation performance at the GenJet level is also analyzed. 
Figure~\ref{cGzwHerror} shows the distributions of average di-jet mass and has an overlapping ratio of $52.6\% \pm 0.25\%$.   
Comparing to the RecoJet distributions, Figure~\ref{cGzwHerror} exhibits much narrow peaks and similar tails.    
That's to say, the peak width of the RecoJet distributions are mainly dominated by the detector performance.             
The correlation between $M_{12}$ versus $M_{34}$ with the GenJets is shown in Figure~\ref{cGzwDiserror}.
Aside from two clearly separable peaks, Figure~\ref{cGzwDiserror} also has a plateau with similar contour and area comparing to Figure~\ref{czwDiserror}, the distribution at RecoJet level.
Clearly, the common patterns of the GenJet and the RecoJet level distributions are induced by the intrinsic boson mass and the jet confusion.

\begin{figure} 
\centering
\includegraphics[width=\columnwidth]{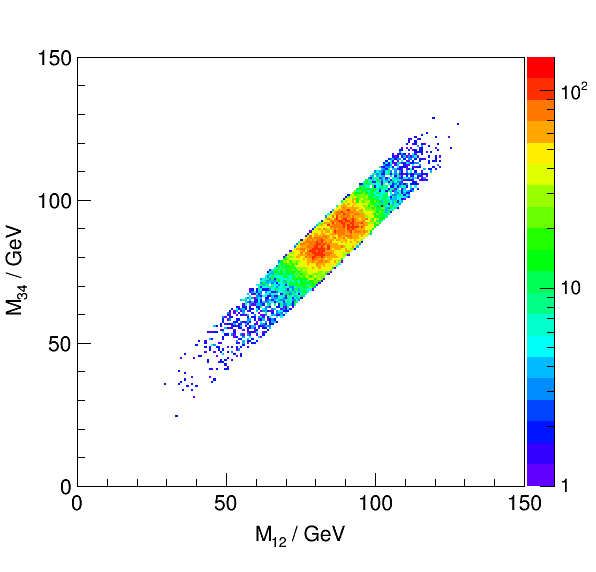}
\caption{ The RecoJet level distribution of $M_{12}$ versus $M_{34}$ with the equal mass condition. The selection efficiency for WW/ZZ is $54\%$/$44\%$.}
\label{czwDissDmass}
\end{figure}

\begin{figure} 
\centering
\includegraphics[width=\columnwidth]{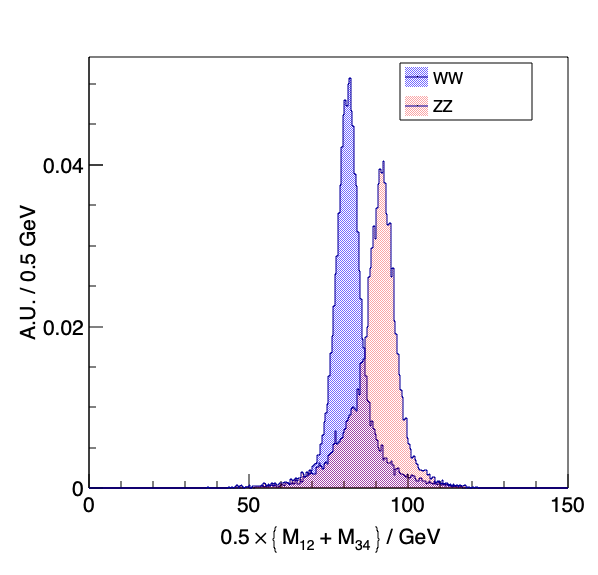}
\caption{ The RecoJet level distributions of $0.5\times(M_{12} + M_{34})$ with the equal mass condition, the overlapping ratio is $39.9\% \pm 0.40\%$.}
\label{czwRdMass1}
\end{figure}

\begin{figure} 
\centering
\includegraphics[width=\columnwidth]{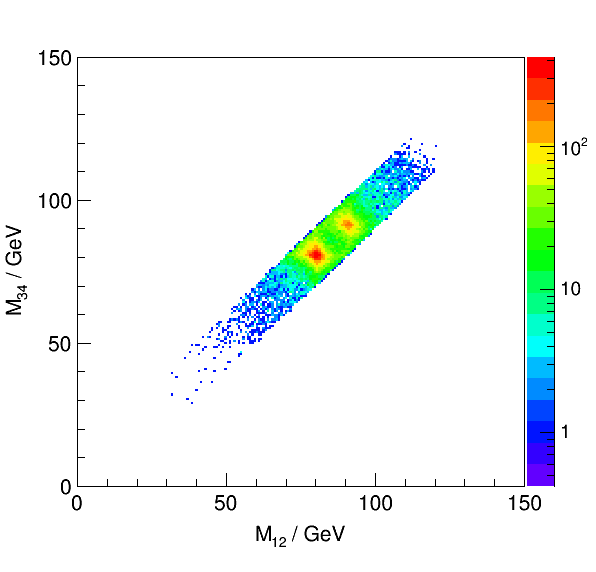}
\caption{ The GenJet level distribution of $M_{12}$ versus $M_{34}$ with the equal mass condition. The selection efficiency for WW/ZZ is $59\%$/$47\%$.}
\label{cGzwDissDmass}
\end{figure}

\begin{figure} 
\centering
\includegraphics[width=\columnwidth]{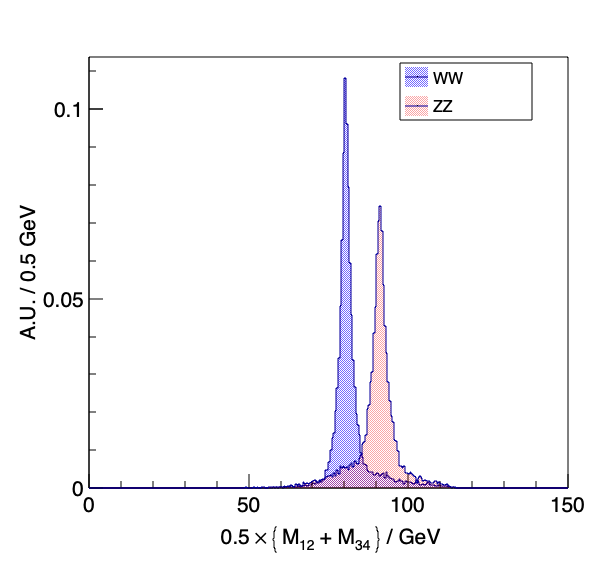}
\caption{ The GenJet level distributions of $0.5\times(M_{12} + M_{34})$ with the equal mass condition, the overlapping ratio is $27.1\% \pm 0.42\%$.}
\label{cGzwRdMass1}
\end{figure}

The area of the plateau can be significantly reduced using the fact that WW and ZZ processes produce two equal mass bosons.
We define an equal mass condition that requires the mass difference between the two di-jet systems to be smaller than 10 GeV ($|M_{12} - M_{34}| < 10$).
This condition keeps roughly half of the events.
After applying this equal mass condition, the overlapping ratios are improved to $39.9\% \pm 0.40\%$ and $27.1\% \pm 0.42\%$, corresponding to the RecoJet and the GenJet plots, 
see Figure~\ref{czwDissDmass} to Figure~\ref{cGzwRdMass1}.

The overlapping ratios of the full hadronic WW and ZZ events can be compared with two reference values. 
The first one is the overlapping ratio at the semi-leptonic di-boson events,
where the invariant mass of the hadronic decayed W and Z bosons can be reconstructed without any jet confusion. 
The second one is the overlapping ratio of the MC truth boson masses, which follow approximately the Breit-Wigner distributions. 
The first value provides a reference to the jet confusion evaluation, 
and the second one describes the impact of intrinsic boson mass distributions and is the lower limit of the overlapping ratio.

 \begin{figure} 
\centering
\includegraphics[width=7.5cm, height = 7.2cm]{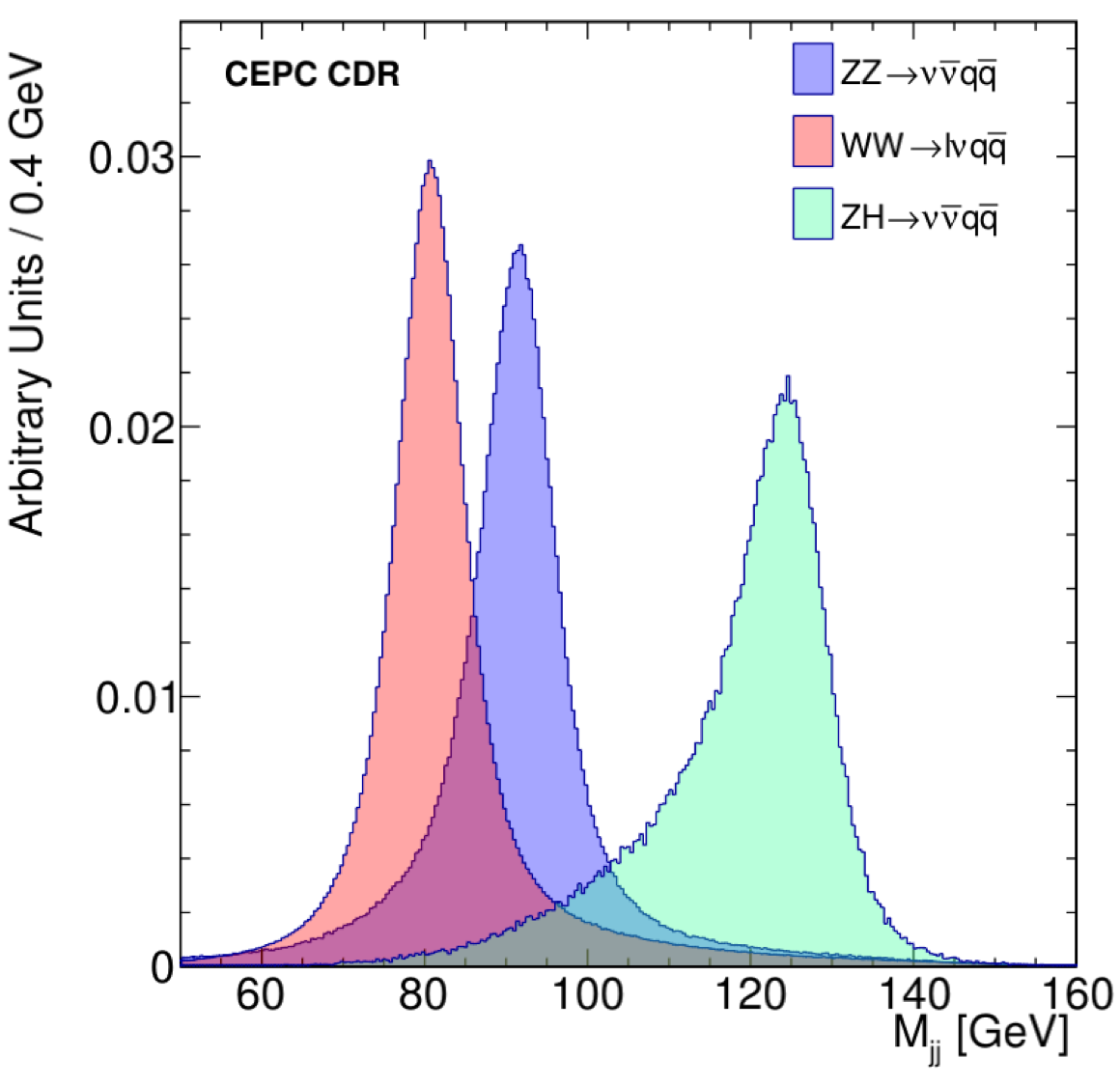}
\caption{The distribution of total invariant mass of hadronic system of $ZZ\to \nu\nu q\bar{q}$, $ZH\to\nu\nu(Z)q\bar{q}(H)$, and $WW\to\mu\nu q\bar{q}$~\cite{r1}. The overlapping ratio of WW and ZZ is $47.32\% \pm 0.26\%$.}
\label{peizhu}
\end{figure}

The invariant hadronic mass distributions of semi-leptonic di-boson events ($ZZ\to \nu\nu qq$, $ZH\to\nu\nu(Z)qq(H)$, and $WW\to\mu\nu qq$, inclusive sample) are shown in Figure~\ref{peizhu}~\cite{r1}.
It has clearly separated peaks at anticipated masses. 
This semi-leptonic overlapping ratio is $47.3\% \pm 0.26\%$.
It is significantly better than that of inclusive full hadronic WW and ZZ events using the RecoJets ($57.8\% \pm 0.23\%$),
but worse than that with equal mass constraint ($39.9\% \pm 0.40\%$).

The overlapping ratios of MC truth boson mass of WW and ZZ events are extracted. 
For the full hadronic events, we calculate the average mass of two MC truth bosons and the overlapping ratio is $13.3\% \pm 0.34\%$. 
For the semi-leptonic event, we extract the truth level value of the mass of the hadronic decay boson, and the overlapping ratio is 12.5\%.
In fact, those two values are close to the integration of two ideal Breit-Wigner distribution overlapping area according to the W and the Z boson masses and widths (12\%).
For simplicity, the average value at full hadronic and semi-leptonic events (12.9\%) is used in later discussion.

Energetic neutrinos can be generated via the semi-leptonic decays at the heavy flavor jet fragmentation, leading to significant missing energy and momentum. 
At the full hadronic WW and ZZ samples, these energetic neutrinos can disturb the jet clustering and pairing performance and increase the jet confusion. 
Its impact is quantified using comparative analysis of the light jet sample. 
Comparing to the inclusive sample, the overlapping ratio at light jet sample is reduced by 7.1\% (from 39.9\% to 32.8\%) and 4.6\% (from 57.8\% to 53.2\%), with and without the equal mass condition respectively.

At 240 GeV center of mass energy, a significant fraction of the WW and ZZ events have energetic ISR photons in their final states.
This ISR effect is included in the Whizard generator. 
These ISR photons, once incident into the ECAL (|cos($\theta$)| < 0.995 at the CEPC baseline), can be recorded as isolated energetic clusters.
Those clusters may also increase of the jet confusion. 
We define an ISR veto condition that excludes events with ISR photons whose energy exceeds 0.1 GeV. 
Once applied on the light jet samples, the overlapping ratio can be further reduced by 3.4\% (from 32.8\% to 29.4\%) and 3.6\% (from 53.2\% to 49.6\%), with and without equal mass condition respectively.

\begin{figure} 
\centering
\includegraphics[width=\columnwidth]{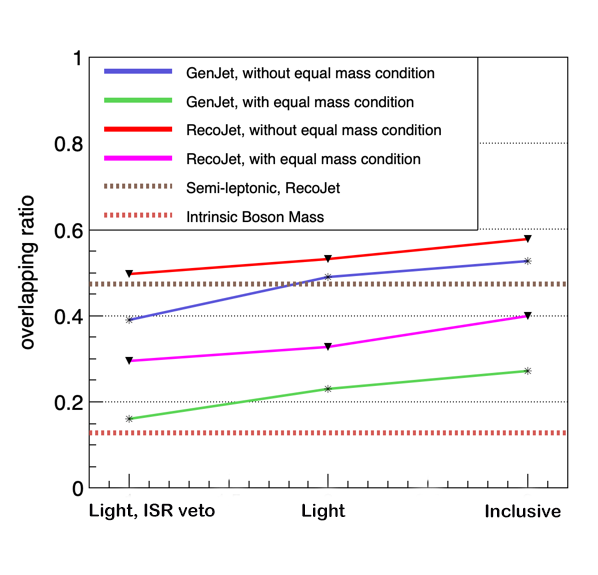}
\caption{ The overlapping ratios for different cases. The X-axis indicates the different sample restrictive conditions: the light flavor samples with ISR veto condition, the light flavor samples, and the inclusive samples. }
\label{cCompare1}
\end{figure}

The same analysis is performed also with GenJets and the overlapping ratio is summarized in Table~\ref{overlappingTable} and Figure~\ref{cCompare1}.
Four lines, corresponding to the cases of the GenJet level or the RecoJet level, with or without the equal mass condition, are identified in Figure~\ref{cCompare1}.
To be compared with two horizontal lines corresponding to the overlapping ratio of truth level boson mass distribution (12.9\%) and that of the semi-leptonic sample (47\%).
Several interesting conclusions can be drawn: 

\begin{itemize}

\item []1, For the full reconstructed samples, the WW and ZZ events could be efficiently separated. 
The separation performance is slightly worse than the semi-leptonic events.
However, the separation performance of the full hadronic events can exceed that of the semi-leptonic events, once the equal mass condition is applied. 

\vspace*{0.3cm}

\item []2, It's actually the jet confusion that dominants the separation performance of the inclusive samples, as the GenJet level samples have already significant overlapping ratio. 
The detector performance is significant on the boson peak width, but contribute only marginally to the overall separation performance. 
For the inclusive samples without equal mass condition, the overlapping ratio only increases by 5\% at the RecoJet level comparing to that at the GenJet level.
Meanwhile, their relative difference becomes more significant once the equal mass condition and other restrictive conditions are applied.

\vspace*{0.3cm}

\item []3, The equal mass condition can efficiently veto events contaminated by large jet confusion. 
At the cost of lost roughly half of the statistic, the separation ratio can be improved by roughly 20\% for both the RecoJets and the Genjets. 
For the GenJets with the light jet samples and ISR photons veto, the overlapping ratio is approaching to the physics lower limit of 12.9\%. 

\vspace*{0.3cm}

\item []4, The neutrinos generated in the heavy flavor jets and the ISR photons contribute approximately a constant amount of overlapping ratio for all four different cases.
In fact, the accumulated impact of neutrinos and ISR photons are larger than that of the detector performance: for the light jet sample with the ISR veto, the RecoJet distribution overlapping ratio ($49.6\% \pm 0.30\%$) is smaller than that of the inclusive sample at the GenJet level ($52.6\% \pm 0.25\%$).
Collectively, they contribute up to 10\% of the overall overlapping ratio on the inclusive sample. 
Therefore, adequate jet flavor tagging and ISR photon finding algorithm can be used, to significantly improve the separation performance. 

\end{itemize}

\begin{table*}
\centering
\caption{The overlapping ratios with different conditions.}
\label{overlappingTable}
\begin{tabular*}{\textwidth}{@{\extracolsep{\fill}}cccc@{}}
\hline
                                             &  Light sample                                           &  Light sample                                                   & Inclusive sample                                          \\
                                             &   non energetic ISR                                  &		&	                                   \\
\hline 
RecoJet	       &   $49.6\% \pm 0.30\%$                            &  $53.2\% \pm 0.29\%$                                     &   $57.8\% \pm 0.23\%$                                 \\
\hline
GenJet                                 &   $39.1\% \pm 0.33\%$                             &  $48.9\% \pm 0.30\%$                                      &   $52.6\% \pm 0.25\%$                                 \\
\hline
RecoJet                                &  \multirow{2}{*}{$29.4\% \pm 0.71\%$ }    & \multirow{2}{*}{$32.8\% \pm 0.49\%$}             &   \multirow{2}{*}{$39.9\% \pm 0.40\%$}         \\
 with equal mass condition   & 		&   		&  	                                     \\
\hline
GenJet                                  &   \multirow{2}{*}{$16.0\%\pm 0.72\%$ }    &  \multirow{2}{*}{$23.0\% \pm 0.51\%$ }            &  \multirow{2}{*}{$27.1\% \pm 0.42\%$}          \\
with equal mass condition    &                                                                   &                                                                          & \\
\hline
\hline
Reference Values	       &						\\
\hline
Semi-leptonic, RecoJet        &                                                                  & $47.3\% \pm 0.26\%$   	 &	                                      \\
\hline
Intrinsic Boson Mass            &                                                                  &  $13.3\% \pm 0.34\%$	 &		\\
\hline
\end{tabular*}
\end{table*}

\section{Quantification of the jet confusion}

In this section, we analyze the correlation between the jet confusion and the overlapping ratio using the angles between the di-jet systems and the MC truth bosons. 
Each event has two di-jet systems and two MC truth level bosons.
The mapping with the minimal value of angle sum is selected. 

\begin{figure} 
\centering
\includegraphics[width=\columnwidth]{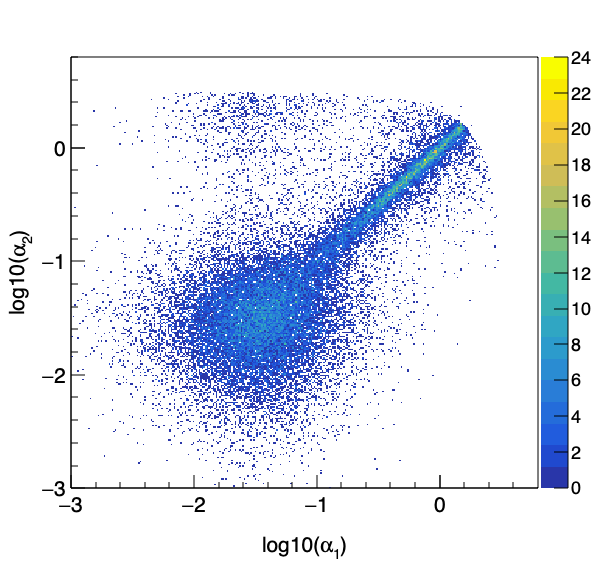}
\caption{The correlation of $\alpha_{1}$ versus $\alpha_{2}$ (unit in radians), the angular difference between reconstructed di-jet systems and the MC truth bosons of the inclusive WW samples.}
\label{cwdR1_dR2}
\end{figure}

Figure~\ref{cwdR1_dR2} shows the correlation of two angles between the RecoJets and the MC truth bosons of the inclusive WW events. 
For $\alpha_{1}$ and $\alpha_{2}$ smaller than 0.1 radians, these two quantities are not correlated.
The distribution actually reflects the jet angle resolution of the CEPC baseline detector.  
For $\alpha_{1}$ and $\alpha_{2}$ larger than 0.1 radians, a strong correlation is observed between these two quantities, corresponding to significant jet confusion. 

\begin{figure} 
\centering
\includegraphics[width=\columnwidth]{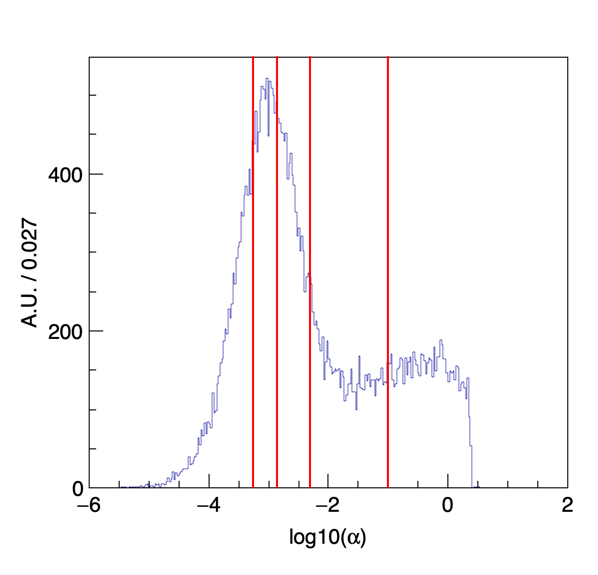}
\caption{The distribution of $\alpha$ ($\alpha = \alpha_{1}\times \alpha_{2}$) of the WW sample using the RecoJets. There are four vertical lines to characterize $\alpha$ into five subsamples, each contains $20\%$ of the statistics.}
\label{cwdR}
\end{figure}

We quantify the jet confusion using the product $\alpha = \alpha_{1}\times \alpha_{2}$ as the order parameter, which increases with the jet confusion. 
Figure~\ref{cwdR} shows the distribution of $Log_{10}(\alpha)$ at the RecoJet level, which exhibits a gaussian-like distribution up to $Log_{10}(\alpha) = -2$ and a flat plateau up to $Log_{10}(\alpha) = 0.4$.
The plateau corresponds to the physics events with large jet confusion. 

To quantify the impact of jet clustering performance, the reconstructed WW sample is divided into five subsamples with the equal statistics, see Figure~\ref{cwdR}. 
A set of thresholds on $\alpha$ are extracted. 
The ZZ samples are divided also into five subsamples using the same thresholds, 
and the overlapping ratios of the same set of subsamples are calculated. 

\begin{figure} 
\centering
\includegraphics[width=\columnwidth]{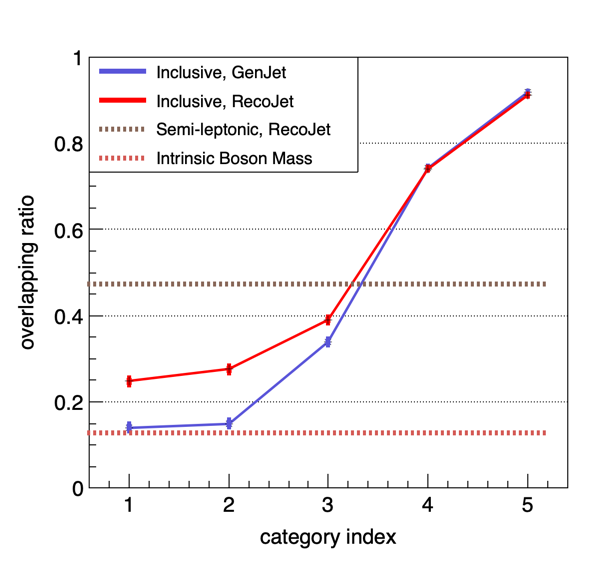}
\caption{ The overlapping ratios of different sets sorted according to the jet confusion order parameter $\alpha$. The red/blue lines is corresponding to the GenJet/RecoJet. The red/brown dashed horizontal line indicates the overlapping ratio of the semi-leptonic sample/intrinsic boson mass distributions, respectively.
}
\label{cGoverlap}
\end{figure}

\begin{figure*}
\begin{minipage}{\textwidth}
\centering
\includegraphics[width=0.19\textwidth]{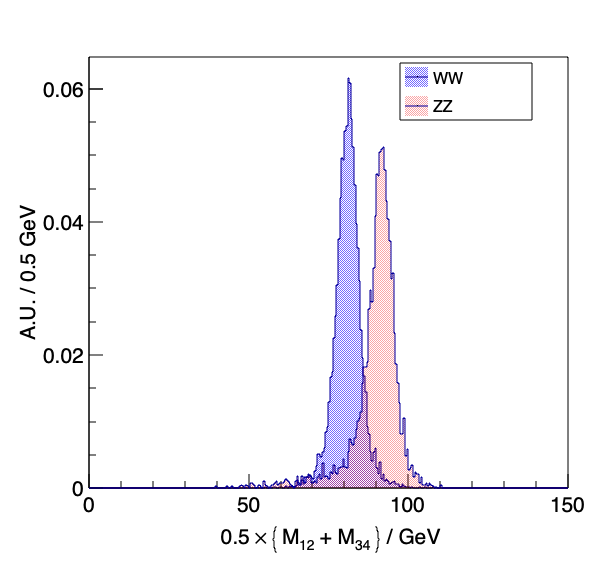}
\hfill
\includegraphics[width=0.19\textwidth]{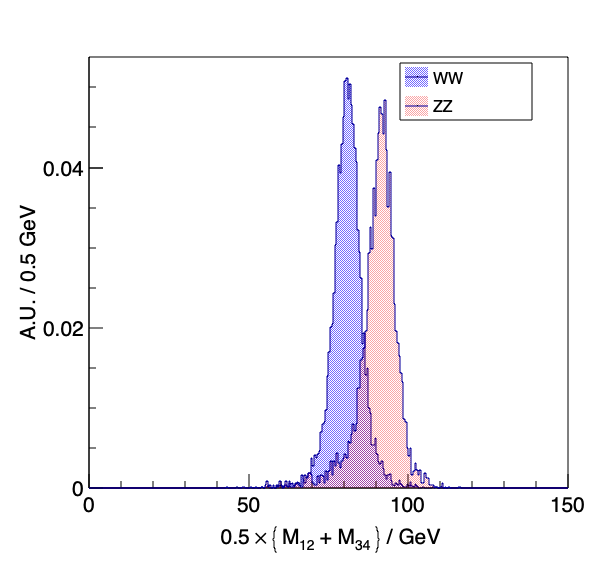}
\hfill
\includegraphics[width=0.19\textwidth]{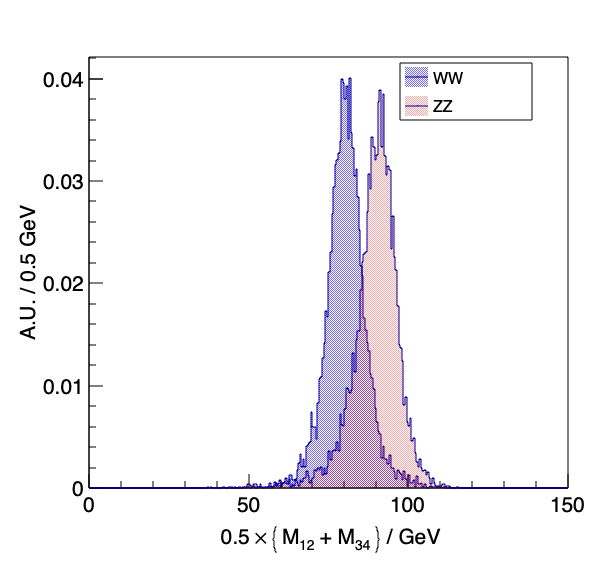}
\hfill
\includegraphics[width=0.19\textwidth]{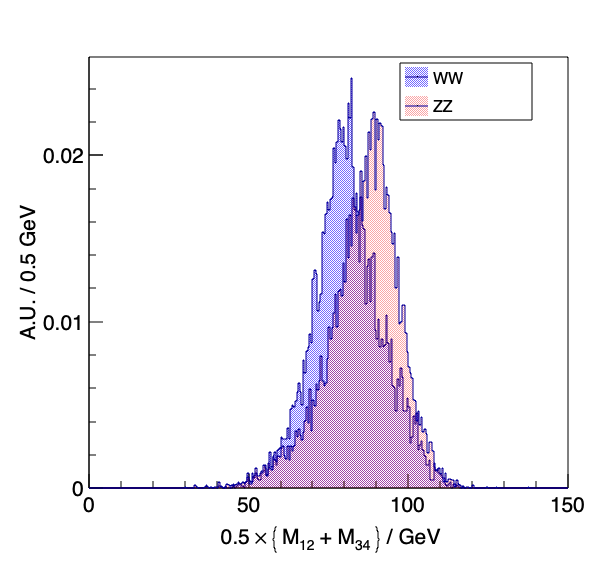}
\hfill
\includegraphics[width=0.19\textwidth]{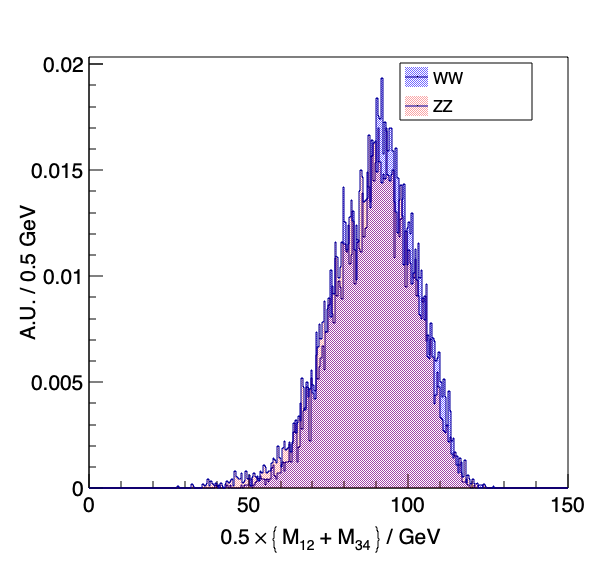}
\\
\includegraphics[width=0.19\textwidth]{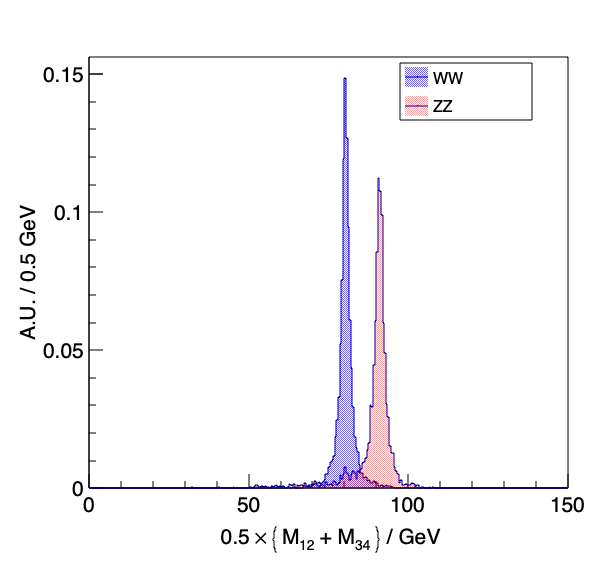}
\hfill
\includegraphics[width=0.19\textwidth]{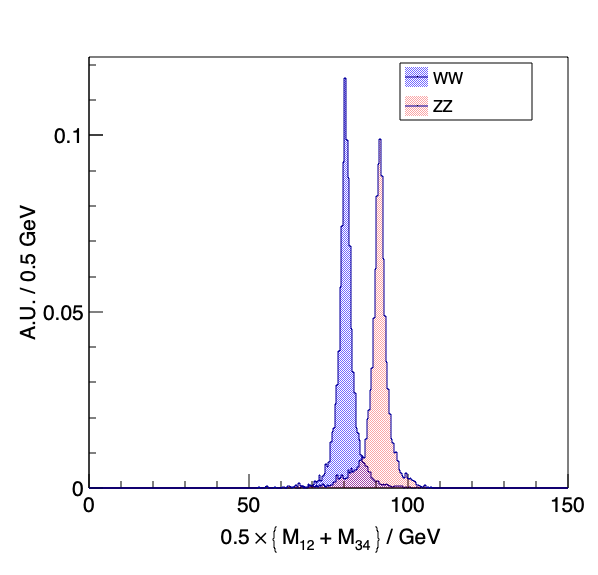}
\hfill
\includegraphics[width=0.19\textwidth]{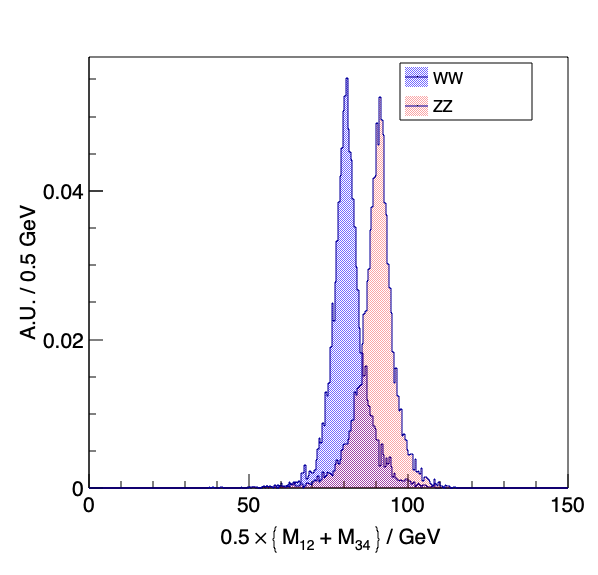}
\hfill
\includegraphics[width=0.19\textwidth]{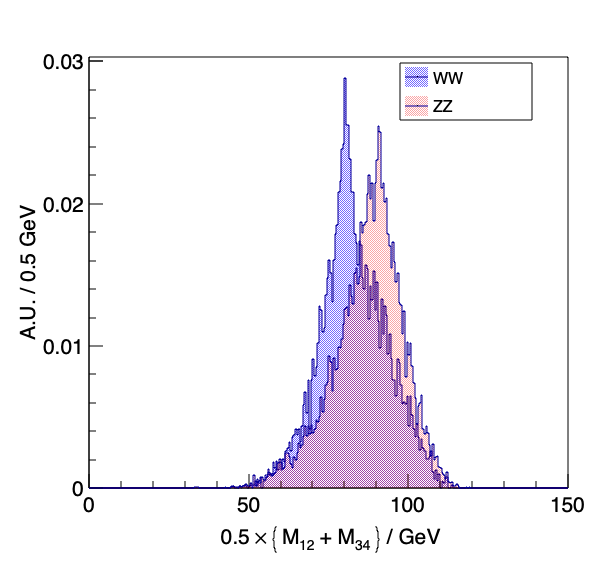}
\hfill
\includegraphics[width=0.19\textwidth]{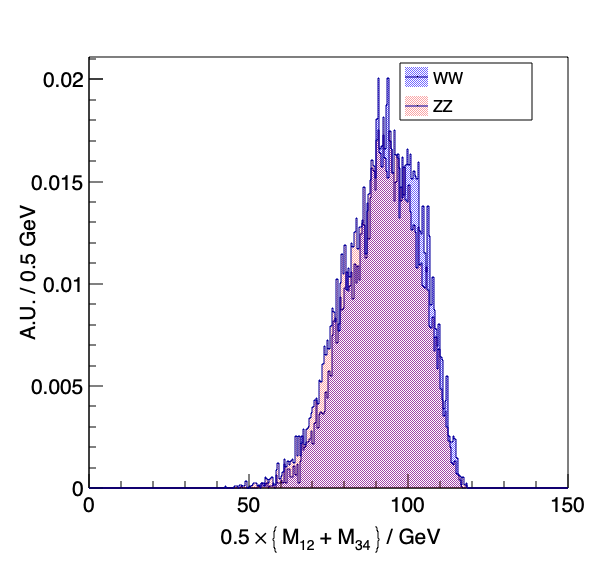}
\end{minipage}
\caption{The average dijet mass distributions after dividing the inclusive sample into five subsamples. From left to right, the $\alpha$ is degrading. The distributions in the top row are using the RecoJets, the overlapping ratio is $24.8\% \pm 0.81\%$, $27.6\% \pm 0.77\%$, $39.1\% \pm 0.63\%$, $74.1\% \pm 0.37\%$ and $91.1\% \pm 0.22\%$, respectively. The bottom distributions are corresponding to the GenJets, the overlapping ratio is $14.1\% \pm 0.89\%$, $15.0\% \pm 0.83\%$, $34.0\% \pm 0.65\%$, $74.4\% \pm 0.37\%$ and $91.9\% \pm 0.21\%$, respectively.}
\label{overlappingRatio}
\end{figure*}

Figure~\ref{overlappingRatio} shows the average di-jet mass distributions of each set at the RecoJet and the GenJet level.
Their overlapping ratios increase monotonically with the jet confusion, see Figure~\ref{cGoverlap}.  
The relative difference between that of the GenJets and the RecoJets, which reflects the detector performance, became less significant. 
In the first set - corresponding to 20\% of the total statistics with the minimal jet confusion, the overlapping ratio of the GenJets is close to the lower limit, and that of the the RecoJets is relatively 76\% larger (14.1\% to 24.8\%). 
In the last set, for both GenJets and RecoJets, the distributions of the WW and ZZ events are similar. 
That's to say, the jet confusion eliminates almost completely the separation power for the last 20\% of statistics with the worst jet confusion. 

 It's interesting that the jet confusion takes polarized pattern in this analysis. 
 Sorting the inclusive samples with the jet confusion, 
 the first 40\% of the samples have only marginal jet confusion (as the overlapping ratio is close to the lower limit).
 However, the jet confusion soon grows to be the leading impact factor of WW/ZZ separation, and dominate the overlapping ratio for the last 40\% of the samples. 
 The critical point occurs at roughly half of the statistics. 
 This S-curve in Figure~\ref{cGoverlap} may characterize profoundly the jet clustering and pairing performance, and can be used as a reference for corresponding performance evaluation and algorithm development.

\section{Conclusion}
\label{secConclusion}

The separation of the full hadronic WW and ZZ events is an important benchmark for the CEPC detector design and performance evaluation. 
This separation performance is determined by the intrinsic boson mass distribution, the detector performance, and the jet confusion. 
Using the CEPC baseline simulation tool, we analyze this benchmark performance using full simulated samples. 
The $\it{e^+e^- k_t}$ and the minimal $\chi^{2}$ methods are used as the jet clustering and pairing algorithms, respectively. 

We quantify the separation performance using the overlapping ratio. 
Comparative analyses are performed to disentangle the impacts of three components. 
The impact of the intrinsic boson mass distribution is characterized by the overlapping ratio of the MC truth boson mass distributions, which is found to be 12.9\%.
The overlapping ratio using the GenJets only includes the intrinsic boson mass and the jet confusion. 
Therefore, the relative difference between the overlapping ratios of the GenJets and the RecoJets describes the impact of detector performance. 
The reconstructed boson masses with hadronic decay final states of the semi-leptonic events are free of the jet confusion. 
These semi-leptonic distributions have an overlapping ratio of $47.3\% \pm 0.26\%$, providing another reference. 

We confirm that the full hadronic WW and ZZ events can be clearly separated at the full reconstruction level.
Using the RecoJets, the overlapping ratio for the inclusive full hadronic WW and ZZ event samples at the CEPC is $57.8\% \pm 0.23\%$. 
An equal mass condition can reduce the overlapping ratio to $39.9\% \pm 0.40\%$, at the cost of vetoing half of the statistics. 
The overlapping ratios of the GenJet level distributions are $52.6\% \pm 0.25\%$ and $27.1\% \pm 0.42\%$, with and without the equal mass condition respectively. 
Comparing to the separation performance with the RecoJets, the GenJets separation performance are significantly improved - especially with the equal mass condition, but its overlapping ratio is still two times larger the lower limit of 12.9\%.
Therefore, we conclude that the jet confusion plays a dominant role in the WW-ZZ separation with full hadronic final states. 

The overlapping ratio for WW and ZZ events with the semi-leptonic final state is estimated to be $47.3\% \pm 0.26\%$, which is between that of the inclusive full hadronic samples with and without equal mass condition ($57.8\% \pm 0.23\%$ and $39.9\% \pm 0.40\%$).
Once the jet confusion is under control, the separation performance of the full hadronic events is better than that of semi-leptonic events, since the former can use mass information from both reconstructed bosons with indepentdent detector response. 

The neutrinos and ISR photons play an important role in the separation performance. 
Collectively, they contribute to roughly 10\% of the overall overlapping ratio. 
Therefore, the jet flavor tagging algorithm and ISR photon identification algorithm are crucial for the full hadronic WW and ZZ event separation. 

The jet confusion is further characterized by the reconstructed angle of bosons. 
The full hadronic WW and ZZ samples are divided into subsamples and sorted accordingly. 
For those subsamples, the jet confusion takes a polarized pattern. 
For the best 40\% of the events, the difference between the reconstructed boson angle and the truth value is smaller than 0.1 radians, 
and the jet confusion is minimum.  
The overlapping ratio of the GenJet level distributions is close to the lower limit of 12.9\%. 
The separation of those events are mainly dominated by the detector performance. 
For the last 40\% of events, the jet confusion dominates the separation performance.

To conclude, our analysis confirms that the baseline CEPC detector and reconstruction software could efficiently separate the full hadronic WW and ZZ events at full reconstruction level. 
The overall separation performance is dominated by the jet confusion. 
Dedicated studies and developments on the jet clustering and pairing algorithms are required, to significantly improve the separation performance.
Adequate ISR photon finding and jet flavor tagging could significantly improve this separation performance. 
Through optimization of the jet clustering and pairing algorithms, for example using differential jet energy resolutions for the jet pairing $chi_2$ calculation, the iterative jet clustering, and the Multiple Variable Analyses, this performance is expected to be improved significantly.

The reconstruction of multi-jets events at the electron positron Higgs factories is critical for the physics reach.
On top of the particle flow reconstruction that produces all the final state particle, 
the critical requirement is to identify precisely all the decay products from each color -singlet. 
In our analysis, the identification is implemented with a straightforward jet clustering and jet pairing algorithm. 
Because the jets at the CEPC can have low energies and large opening angles, these algorithms can lead to large jet confusions that dominate the final measurement accuracy. 
Dedicated studies to control the jet confusion, or equivalently, the development of color-singlet reconstruct algorithms, are critical.
The WW/ZZ separation analysis presented in this paper is an early step of these studies. 
It not only demonstrates the physics performance of the CEPC baseline but also provides the reference and a simple quantification method to evaluate different color-singlet reconstruction algorithms.

\begin{acknowledgements}
We are in debt to Jianming Qian, Liantao Wang, Huaxing Zhu, and Haibo Li for their constructive suggestions.
We are grateful to Dan Yu, Xianghu Zhao, Hao Liang, and Yuxuan Zhang for their supports and helps.
We thank Gang Li and Chengdong Fu for producing the samples. 

This work was supported by National Key Program for S\&T Research and Development (Grant No.: 2016YFA0400400), the National Natural Science Foundation of China (Grant No.: 11675202), the Hundred Talent Programs of Chinese Academy of Science (Grant No.: Y3515540U1).
\end{acknowledgements}


\begin{thebibliography}{4}

\bibitem{r1}
The CEPC Study Group, CEPC Conceptual Design Report: Volume 2 - Physics \& Detector, arXiv:1811.10545 [hep-ex]

\bibitem{r6}
M.Q. Ruan, Reconstruction of physics objects at the Circular Electron Positron Collider with Arbor, Eur. Phys. J. C \textbf{78}, 426 (2018) no.5

\bibitem{ILCTDR}
T. Behnke et al.,The International Linear ColliderTechnical Design Report-Volume 4: Detectors (2013). ArXiv: 1306.6329

\bibitem{CLICCDR}
The CLIC Collaboration, CLIC Conceptual Design Report (2012). CERN-2012-007

\bibitem{r2}
W. Kilian, T. Ohl, J. Reuter, WHIZARD: Simulating Multi-Particle Processes at LHC and ILC , Eur. Phys. J. C \textbf{71}, 1742 (2011) 

\bibitem{r3}
M. Moretti, T. Ohl, J. Reuter, O'Mega: An Optimizing matrix element generator , LC-TOOL-2001-040-rev, arXiv: hep-ph/0102195-rev

\bibitem{r4}
The Pythia Group, An Introduction to PYTHIA 8.2, Comput. Phys. Commun. 191 (2015) 159-177, arXiv:1410.3012 [hep-ph]

\bibitem{r8}
C.D. Fu, Full Simulation Software at CEPC, \url{http://cepcdoc.ihep.ac.cn/DocDB/0001/000167/001}. Accessed 23 Oct 2017

\bibitem{r5}
The GEANT4 Collaboration, S. Agostinelli et al., GEANT4: A Simulation toolkit, Nucl. Instrum. Meth. A506 (2003) 250-303

\bibitem{r7}
M. Cacciari, G.P. Salam and G. Soyez, Eur. Phys. J. C \textbf{72}, 1896 (2012)

\bibitem{r9}
M. Tanabashi et al. (Particle Data Group), Phys. Rev. D 98, 030001 (2018)

\bibitem{r12}
M.Q. Ruan, Druid: event display for the linear collider, arXiv:1303.3759 [physics.ins-det]


\end{thebibliography}
\end{document}